 \documentclass[reviewcopy]{elsart}

\usepackage{natbib}

\usepackage{graphicx}

\usepackage{amssymb}

\usepackage{lineno}
\linenumbers

\begin{document}

\begin{frontmatter}



\title{Reliability of resistivity quantification for shallow subsurface water processes}


\author[t1]{J.~Rings\corauthref{c1}}
\corauth[c1]{Corresponding author. Address now: Joerg Rings, ICG-IV (Agrosphere), Institute of Chemistry and Dynamics of the Geosphere, Forschungszentrum Juelich GmbH, 52425 Juelich, Germany}
\ead{j.rings@fz-juelich.de}
\author[t1]{C.~Hauck}

\address[t1]{Institute for Meteorology and Climate Research\\Karlsruhe Institute of Technology}

\begin{abstract}

The reliability of surface-based electrical resistivity tomography (ERT) for quantifying resistivities for shallow subsurface water processes is analysed. A method comprising numerical simulations of water movement in soil and forward-inverse modeling of ERT surveys for two synthetic data sets is presented. Resistivity contrast, e.g. by changing water content, is shown to have large influence on the resistivity quantification.\\
An ensemble and clustering approach is introduced in which ensembles of 50 different inversion models for one data set are created by randomly varying the parameters for a regularisation based inversion routine. The ensemble members are sorted into five clusters of similar models and the mean model for each cluster is computed. Distinguishing persisting features in the mean models from singular artifacts in individual tomograms can improve the interpretation of inversion results.\\
Especially in presence of large resistivity contrasts in high sensitivity areas, the quantification of resistivities can be unreliable. The ensemble approach shows that this is an inherent problem present for all models inverted with the regularisation based routine. The results also suggest that the combination of hydrological and electrical modeling might lead to better results.

\end{abstract}


\end{frontmatter}


\section{Introduction}
The quantification of water content by geophysical methods is an important focus of hydrogeophysical research. Surface based electrical resistivity tomography (ERT) is a promising method, because it is non-intrusive and can cover large surface areas quickly, while it might also be permanently installed for automated monitoring purposes. The development of inversion software for the processing of measured (apparent) resistivities to models of true resistivity has made fast and extensive surveys possible \citep{daily04}. Consequently, assessing the reliability of ERT for quantifying soil water content is a currently active research field.\\
ERT has successfully been used in a number of different applications, e.g. in borehole surveys of tracer experiments \citep{slater00,kemna02} or in laboratory experiments \citep{binley96,slater02}. It has also been applied in surface-based surveys of the vadose zone \citep[e.g.][]{daily92} and of groundwater flow after heavy rain \citep{suzuki01}.\\
Because choice of measurement configuration and inversion parameters may have significant influence on the survey results, improving the quality of ERT surveys has been an intense research topic. \citet{dahlin04} have compared 10 different electrode arrays for 2D surveys and assessed their quality using synthetic data sets. \citet{stummer04} have developed algorithms to calculate optimal electrode arrays that provide as much information on the subsurface as possible. The effects of measurement errors \citep{zhou03,oldenborger05} and geometry \citep{loke00,hennig05,sjoedahl06} and inversion parameters \citep{carle99, rings08} on the surveys have been studied.\\
Geophysical methods cannot directly determine hydrological properties like soil water content. They must be deducted using a general or calibrated relationship between the attribute of interest and the property available through geophysical measurements. In the case of ERT, the resistivities of the subsurface are related to water content by a generic petrophysical relation; usually the equation by \citet{archie42}. The resistivities, again, are not readily available from surface-based ERT surveys, but must be obtained from the measured apparent resistivities via inversion. The most widespread inversion methods rely on regularised least-squares minimisation to find the smoothest model of resistivities that gives a model response closest to the measured apparent resistivities.\\
Even assuming that the petrophysical relation between resistivity and water content is known, the resistivity models are non-unique and have likely been affected by the inversion process. The sensitivity of tomographic surveys plays a major role in the retrieval of subsurface characteristics, e.g. for surface-based ERT the sensitivity decreases with depth. Low sensitivity areas (but not only those) can often be plagued by inversion artifacts \citep[e.g.][]{rings08}. The inversion process and the choice of inversion parameters, e.g. the regularisation parameters, determine how well the inverted model will reproduce the real distribution. However, some of the parameter choices can not reliably be based upon observation, but must be fitted or depend on experience.\\
\citet{daylewis04,daylewis05} refer to the loss of information caused by the inversion process, lack of sufficient prior information and survey geometry as 'correlation loss'. They developed a method to compute the correlation loss as a function of the influencing factors. This allows an analytical integration of these factors into geostatistical analyses of quantitative hydrological field surveys, but needs a priori knowledge of covariance models. \citet{singha06b} suggest a nonstationary estimation approach that uses numerical simulations of transport and electrical current flow to deduct apparent petrophysical relations. These methods modify the translation from the inverted models by adjusting the petrophysical relation but require either a priori knowledge or are computationally intensive.\\
To assess the quality of ERT-based water content quantification, the complete processing chain including the inversion process, the petrophysical relation and numerical simulations of the soil water movement has to be evaluated. This study introduces a combined approach using soil hydraulic simulations and ensemble building of inverted models to estimate the uncertainty inherent in typical applications of ERT for water content quantification.
\section{Methods}
To evaluate the inversion process, a forward-inverse cycle approach is used. In numerous applications and studies, forward modeling of synthetic data sets has been used to gain additional insight and confidence into measurements and the inversion process \citep[e.g.][]{loke02b, godio03, hauck03, loke03, nguyen05, nguyen07, rings05}. Forward modeling routines are applied to synthetic data sets obtained from simulations of soil water movement. For two cases studies, the approach is used to discuss how slight variations in the soil structure influence the resistivity retrieval, and thereby the water content retrieval.\\
The second part of the study proposes an ensemble approach which allows an overview of the possible range of inverted models, improves the analysis and enables general assertions about how well a given model can be characterised through the chosen inversion process.\\
In the following, each methodological step of the methods will be shortly introduced, further discussion will illustrate how these steps can be applied to create and analyse two synthetic data sets.\\
The forward-inverse cycle consists of three steps:
\begin{enumerate}
\item \emph{Simulation of water movement in soil}: A model with specific soil structure is generated for numerical simulation of water movement. The movement of a water front, caused by infiltrating rainfall, is simulated over time. Characteristic states of water percolation are identified (starting with a completely dry soil) and a simplified distribution of water content for each state is extracted. 
\item \emph{Generic resistivity model}: A generic resistivity model mirroring the soil structure from (1) is created. 
\begin{itemize}
\item For a model representing a dry state (no water content), resistivities are assigned based on typical values known from laboratory measurements and/or literature.
\item For states of water percolation, changes in water content can be calculated using the water content distribution from (1). They can be transferred into resistivity changes by applying a petrophysical relation, e.g. the equation by \citet{archie42}.
\item A finite-element based forward modeling routine transfers the generic resistivity models into model responses (sets of apparent resistivities) that correspond to the data that would have been recorded by field surveys. Random noise is added to simulate field measuring conditions.
\end{itemize}
\item \emph{Resistivity inversion}: The apparent resistivities are inverted using a suitable inversion scheme. The most widespread inversion schemes include smoothness constrained (L2-norm) methods and robust (L1-norm) schemes which are preferable if sharp layer boundaries are present. The forward-inverse cycle is completed by comparing and evaluating the generic and inverted model of resistivities.
\end{enumerate}
The ensemble method comprises two steps:
\begin{enumerate}
\item \emph{Ensemble generation}: For each data set, an ensemble of 50 different inverted models is created by varying the inversion parameters and/or the inversion scheme. The parameter set is chosen randomly from a parameter space constrained to physically meaningful parameter sets.
\item \emph{Clustering}: A clustering algorithm is used to group similar models of the ensemble. Cluster members can be averaged to simplify the analysis of the ensemble.
\end{enumerate}
\subsection{Forward-inverse cycle}
The application of this methodology was governed by the available software codes for modeling and inversion. This section discusses how the steps were specifically realised to create and analyse two synthetic data sets.
\subsubsection{Simulation of water movement in soil}
A numerical simulation of water movement was used to ensure that realistic distributions of water content (and thus resistivity) were used in this study.\\
If a continuously connected air phase is assumed, the equation of motion for water in soil was given by \citet{richards31} as:
\begin{equation}\label{eq:richards}
\frac{\partial}{\partial t} \theta_w + \nabla \cdot [ K_w (\nabla \Psi_m - \varrho_w\vec{g})]
\end{equation}
with volumetric water content $\theta_w$, hydraulic conductivity $K_w$, matric potential $\Psi_m$, density of water $\varrho_w$ and gravitational acceleration $\vec{g}$. To solve Eq. \ref{eq:richards} for water content, the material properties have to be given that connect $\theta_w$, $K_w$ and $\Psi_m$. Usually, the soil-water characteristic $\theta_w(\Psi_m)$ and the conductivity $K_w(\theta_w)$ are parameterised.\\
The most widely used parameterisation for the soil-water characteristics \citep{vangenuchten80}, written in terms of water saturation $S = (\theta - \theta_r)/(\theta_s - \theta_r)$ with residual water content $\theta_r$, saturated volumetric water content $\theta_s$ and hydraulic head $h_m = \Psi_m/(\varrho_w g)$, is
\begin{equation}
S(h_m) = [1+ (\alpha h_m)^\nu]^{-1+\frac{1}{\nu}}
\end{equation}
with the scaling factor $\alpha$, which is related to the air-entry value $1/\alpha$, and the parameter $\nu$ connected to the pore size distribution. The hydraulic conductivity is characterised by applying the parameterisations of \citet{mualem76}. A concise overview of the soil physics is given e.g. by \citet{stephens96}.\\
Equation \ref{eq:richards} was solved numerically using the HYDRUS software \citep{simunek06}. By defining time-variable precipitation and evaporation rates as atmospheric boundary conditions, changes in the hydraulic head $h_m$ and thus water movement are induced.\\
The simulations were conducted with models representing a two-layered soil representative of a site used in previous field studies \citep{rings08}. In addition to an atmospheric boundary, a seepage boundary on the bottom allowed water to leave the domain. From the simulations, characteristic states of a water front infiltrating the domain were identified. Generally, beyond the dry state, characteristic states should be chosen at times when the water content distribution has changed significantly, e.g. when one layer has become completely saturated.\\

\subsubsection{Generic resistivity model}
The transfer from water saturation values $S$ to electrical resistivity $\rho$ is given by the equations of \citet{archie42}. Here the quotient form is applied given by
\begin{equation}\label{eq:archie}
\frac{\rho_i}{\rho_j}=\left(\frac{S_j}{S_i}\right)^{-n}
\end{equation}
where it is assumed that two measurements of the same soil at time steps $i,~j$ differing only in water saturation are connected by the saturation exponent $n$. $n$ is near 2 for an organic overburden and in the range of 1.01 to 2.7 for unconsolidated sands \citep{ulrich04}.\\
A generic model of resistivities was constructed calculate the response (measurement data) an actual ERT survey would have retrieved. We simulated field conditions by superimposing 3 \% random noise on the resulting apparent resistivity data set.

\subsubsection{Inversion of apparent resistivities}
The forward-inverse cycle is completed by inverting the simulated measurement data. Generic and inverted models can then be compared and the discrepancies analysed.\\
A robust inversion scheme by \citet{loke03}, which is usually employed whereever sharp layer boundaries are expected, was chosen. It is implemented as an iteratively reweighted least-squares method \citep{wolke88} in the software RES2DINV:
\begin{equation}\label{eq:robustpap}
(\mathbf{J}_i^T\mathbf{R}_d\mathbf{J}_i+\lambda_i\mathbf{W}^T\mathbf{R}_m\mathbf{W})\Delta\mathbf{m}_i=\mathbf{J}_i^T\mathbf{R}_d\Delta\mathbf{d}_i-\lambda_i\mathbf{W}^T\mathbf{R}_m\mathbf{W}\mathbf{m}_{i-1}
\end{equation}		
Here $\mathbf{J}_i$ are Jacobian matrices of partial derivatives for the $i$-th iteration, $\mathbf{W}$ is a roughness filter using a first-order finite-difference operator \citep{degroot90}, $\lambda_i$ are damping factors, $\mathbf{R}_d$ and $\mathbf{R}_m$ are weighting matrices to give different elements of data misfit and model roughness vectors equal weights, $\Delta\mathbf{m}_i$ is the change in model parameters for the $i$-th iteration and $\Delta\mathbf{d}_i$ is the data misfit vector containing the difference between calculated and observed apparent resistivities. Since the $\Delta\mathbf{d}$ values may extend over several orders of magnitude, logarithmic differences are employed.\\
Equation \ref{eq:robustpap} is solved iteratively until either the root-mean square (RMS) of the data misfit vector $\Delta\mathbf{d}_i$ does not change significantly after an inversion step and/or it becomes smaller than the measurement accuracy. The weighting matrices $\mathbf{R}_d$ and $\mathbf{R}_m$ are predefined, and default values were chosen for $\lambda_i$.
\subsection{Ensembles}
Inversion problems for geoelectrical surveys are usually ill-posed, mixed determined problems. If the errors in data acquisition and in the inversion process would be known quantitatively, the optimum model and its error distribution could be determined exactly. Measurement errors often can only be estimated, and further discrepancies may be introduced during inversion, especially if an inversion code is used that does not rigorously optimise for a given error estimate. Additionally, inverted models can be plagued by possibly large inversion artifacts depending e.g. on resistivity contrasts.
\subsubsection{Building Ensembles}
Consequently, it might not be sufficient to analyse only the optimum model (i.e. the model with the smallest data misfit), but to compute a range of possible models addressing the inherent variability of the inversion process. By randomly varying the inversion parameter set and creating an ensemble of possible inversion models, the whole parameter space and thus the possible model range is explored.\\
For the RES2DINV code used here, the selected parameters are listed in Table \ref{tab:ensemble}. The table also includes for each parameter the range from which a value was automatically and randomly selected. The parameter selection encompasses the use of smoothness constrained and robust inversions as well as two mixed formulations with a robust constraint applied only on the data, and one with a robust constraint applied only on the model. Further variations address the regularisation, e.g. the damping factor, where an initial damping factor $\lambda_{start}$ and the maximum damping factor $\lambda_{max}$ are varied. For most variations, the maximum damping factor $\lambda_{max}$ is kept at $\lambda_{max}=10\cdot \lambda_{start}$. Additional variations include the reduction of side block effects, the ratio of vertical to horizontal smoothness filtering and the use of the first iteration step model as a reference model for the further iterations instead of using the average of resistivities.\\
It should be noted that this choice of variations is specific for the software used in this study. However, the idea can easily be transferred to similar inversion approaches.\\
Almost all inversions resulted in inverted models with RMS errors smaller than 4\% as can be expected from adding 3\% artifical noise to the data set. Some single inversions, however, resulted in a larger RMS error. In section \ref{synthcase}, both, inversion models with RMS $\le 4\%$ and $> 4\%$, will be included to keep the ensembles balanced.
\subsubsection{Clustering}\label{clustering}
Each ensemble is created as a set of 50 different inversion models and then regrouped using a $k$-means clustering algorithm \citep{dubes88}. 50 models have been chosen arbitrarily as a compromise between computational efficiency and the necessity to generate a sufficiently large ensemble for clustering. The $k$-means clustering method starts with a collection of \emph{genes}; here a gene is a row of all block resistivities of one model. The distance $d$ between two genes is calculated as a Pearson correlation
\begin{equation}\label{eq:pearson}
d=\frac{1}{N}\sum \left( \frac{x_i - \overline{x}}{\sigma_x} \right) \left( \frac{y_i - \overline{y}}{\sigma_y} \right)
\end{equation}
where $\overline{x}$ is the average of values in gene $x$ and $\sigma_x$ is the standard deviation of these values \citep{eisen98}. The $k$-means clustering starts with a user decision on the number of clusters to be created, then randomly assigns each gene to a cluster. For each cluster, the average model is created, then each gene is assigned to the cluster it has the smallest distance from. These last steps are repeated until an optimal solution is found. At least two runs creating the same optimal solution are needed to reach a reliable solution \citep{eisen98}. In this study, we used five clusters to generate a sufficiently large cluster variability while ensuring that the number of ensemble members per cluster is not too small.

\subsection{Applicability}
All five steps presented here form an analysis cycle for a synthetic case study investigating the reliability of resistivity quantificaton for shallow subsurface water processes. For application to field cases, it is possible to create and analyse a simplified synthetic representation of the actual site following the five steps above o rapplay only the ensemble and clustering steps t determine the spread of possible inversion results. 

\section{Synthetic case studies}\label{synthcase}
All test cases studied here are based on a simple two layer medium representing the structure of a full-scale dike model described in detail by \citet{rings08}. Although synthetic data sets are employed to distinctively focus on specific anomalies, the material parameters were obtained from real observation. Hydraulic parameters, following the van Genuchten-Mualem parameterisation, were determined in a laboratory experiment by \citet{scheuermann05}. For obtaining the parameters of the overburden, we used an inversion procedure supported by the HYDRUS software. As no direct measurements of water content in the overburden were available, a rainfall experiment described in \citet{scheuermann05} was simulated. Pressure head measurements in the sand, but directly below the overburden, were used to invert the hydraulic parameters of the overburden. The resulting parameters are listed in Table \ref{tab:materials} for the two different materials. Meteorological data from the permanent station Karlsruhe-Nordwest (Germany) were used as forcing. The Penman formula, calibrated for grass cover by \citet{doorenbos77}, was applied to these data to retrieve values for potential evapotranspiration. Combined with measured precipitation rates, these values have been used as daily averages for simulations of 210 days based on measurements in 2001.\\
Based on this two layer medium, two generic cases representing different idealised case studies were created: The first case study simulates a defective sealing, where an infiltration plume of water is generated in the sand layer. In the second case study, a rectangular, hydraulically resistive anomaly is placed in the sand underneath the organic overburden. 
\subsection{First Case: Defective Sealing}
The first case is based on the idea of a crack in a dike sealing. Damaged sealings are critical, as even through small cracks, large amounts of water can infiltrate.\\
In this hypothetical case, water infiltrates through an otherwise sealed off surface through one crack. The sealing is considered to be invisible to the geoelectrical survey.
\subsubsection{Water Simulation}
\begin{figure}
\begin{center}
\includegraphics*[width=\linewidth]{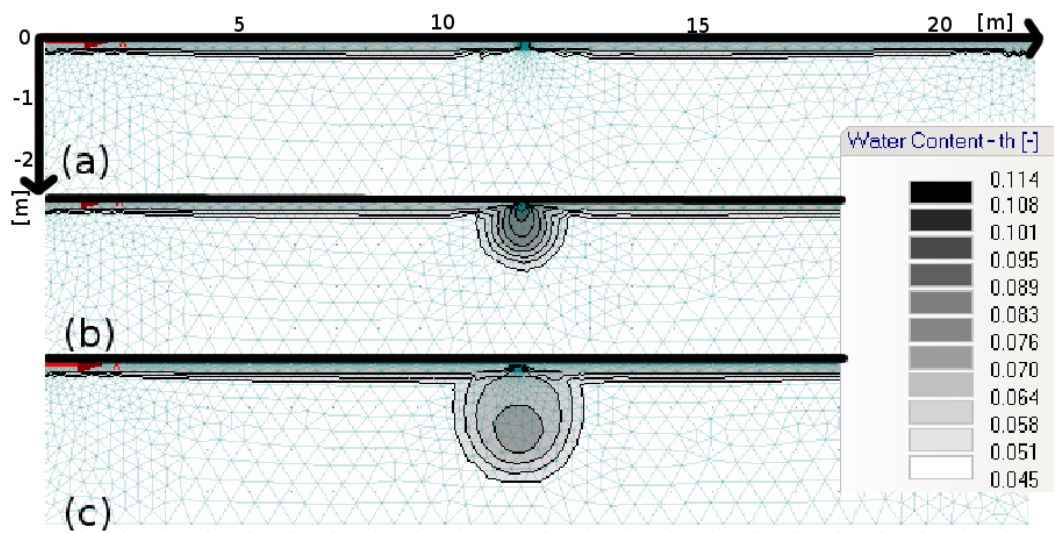}
\end{center}
\caption{\label{fig:seal-states}Defective sealing, characteristic states of water percolation. (a) Dry State (b) Infiltration State (c) Diffusion State.}
\end{figure}

 In HYDRUS, the sealing is modeled as a no-flow boundary, and the crack has an atmospheric boundary and is filled with sand material. The simulation results show water infiltrating through the crack into the sand where it diffuses into a sinking plume. The water content does not change outside of the plume (Fig. \ref{fig:seal-states}). Three characteristic states of the simulated results can be identified: dry state (Fig. \ref{fig:seal-states}a), infiltration state (the plume begins to form in the sand, Fig. \ref{fig:seal-states}b) and the diffusion state (Fig. \ref{fig:seal-states}c), where the center of the plume has propagated into the sand and the top layer is already drying. The transfer from water content to resistivities was done by assuming a dry state resistivity of $\rho=400~\Omega m$ for the overburden and $\rho=5000~\Omega m$ for the sand and applying Eq. \ref{eq:archie} with saturation exponent $n=2$ for the overburden and $n=1.164$ for the sand \citep[see][]{rings08}. During infiltration and diffusion, this results in a minimal resistivity in the plume of $\rho=2000~\Omega m$.\\
\subsubsection{Forward-Inverse Cycle}
\begin{figure}
\begin{center}
\includegraphics[width=0.75\textwidth]{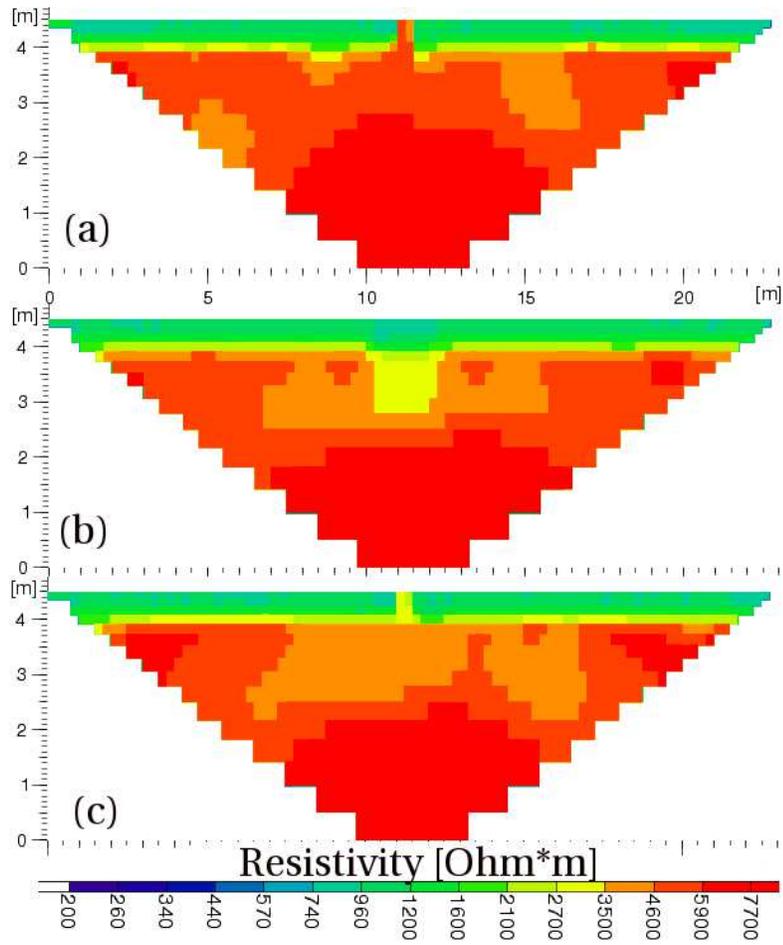}
\end{center}
\caption{\label{fig:seal-inv}Inverted models for the case of the defective sealing. (a) Dry State (b) Infiltration State (c) Diffusion State.}
\end{figure}

\begin{figure}
\begin{center}
       \includegraphics[width=.75\linewidth]{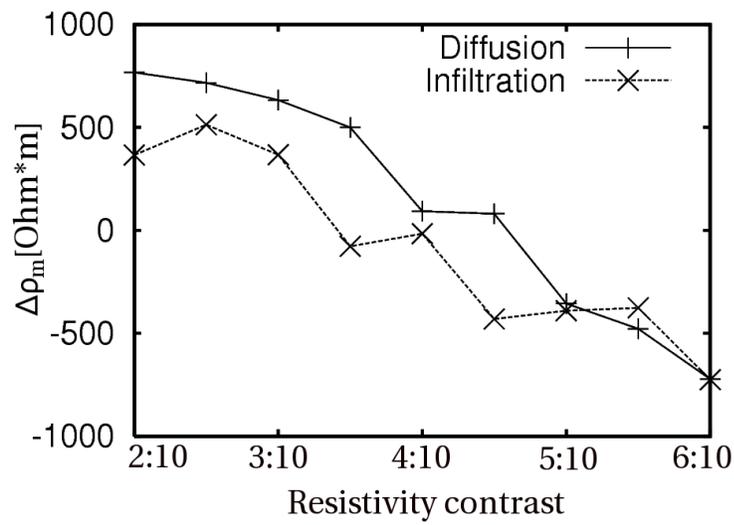}  
\end{center}
\caption{\label{fig:seal-vari}Misfit of the anomaly for the case of the defective sealing, shown is the resistivity contrast as the ratio plume divided by host material (x-axis) versus $\Delta\rho_m$ of the the inverted model (y-axis).}
\end{figure}

Figure \ref{fig:seal-inv} shows three standard (robust) inversion models for the three states of water percolation. A complete Wenner-Schlumberger array with electrode separation 0.5 m has been simulated in the forward modeling. In the dry state, the crack is clearly visible. In the infiltration state, the infiltrating plume is characterised through a distinct lower resistivity than the background, while in the diffusion state the inversion did not sufficiently resolve the shape of the plume.\\
To analyse the dependence of the inversion results on the resistivity contrast between the plume and the host material, the plume resistivity was increased or lowered in steps of $250~\Omega m$ around the minimal plume resistivity of $2000~\Omega m$. A total of nine models with plume resistivity ranging from $1000$ to $3000~\Omega m$ were explored, while the background resistivity stayed constant at $5000~\Omega m$.\\
Generally, the resistivity of an anomaly $\rho_{anom}^{-}$ is
\begin{equation}
\rho_{anom}^-=\min\{ \rho_i \}
\end{equation}
for all model blocks $i$ below the overburden. The misfit in the anomaly's resistivity $\Delta\rho_m$ is the difference between the resistivity of the anomaly in the generic $\rho_{anom,gen}^-$ and inverted model $\rho_{anom,inv}^-$:
\begin{equation}\label{eq:misfit}
\Delta\rho_m= \rho_{anom,gen}^- - \rho_{anom,inv}^-
\end{equation}
For this case study, $\rho_{anom}^-$ corresponds to the resistivity in the center of the plume. Figure \ref{fig:seal-vari} shows the results of the forward-inverse cycle as $\Delta \rho_m$ vs the resistivity contrast. While the error in resistivity quantification is smallest for the orginal contrast of 4:10, smaller and higher contrasts both result in increasingly larger $\Delta\rho_m$. \\
$\Delta\rho_m$ is slightly smaller in the infiltration state. In the diffusion state, the center of the plume has sunk to greater depth, where the reduced sensitivity of ERT may be the reason for a less accurate quantification. 
\subsubsection{Ensemble}
\begin{figure}
\begin{center}
\includegraphics*[width=.89\linewidth]{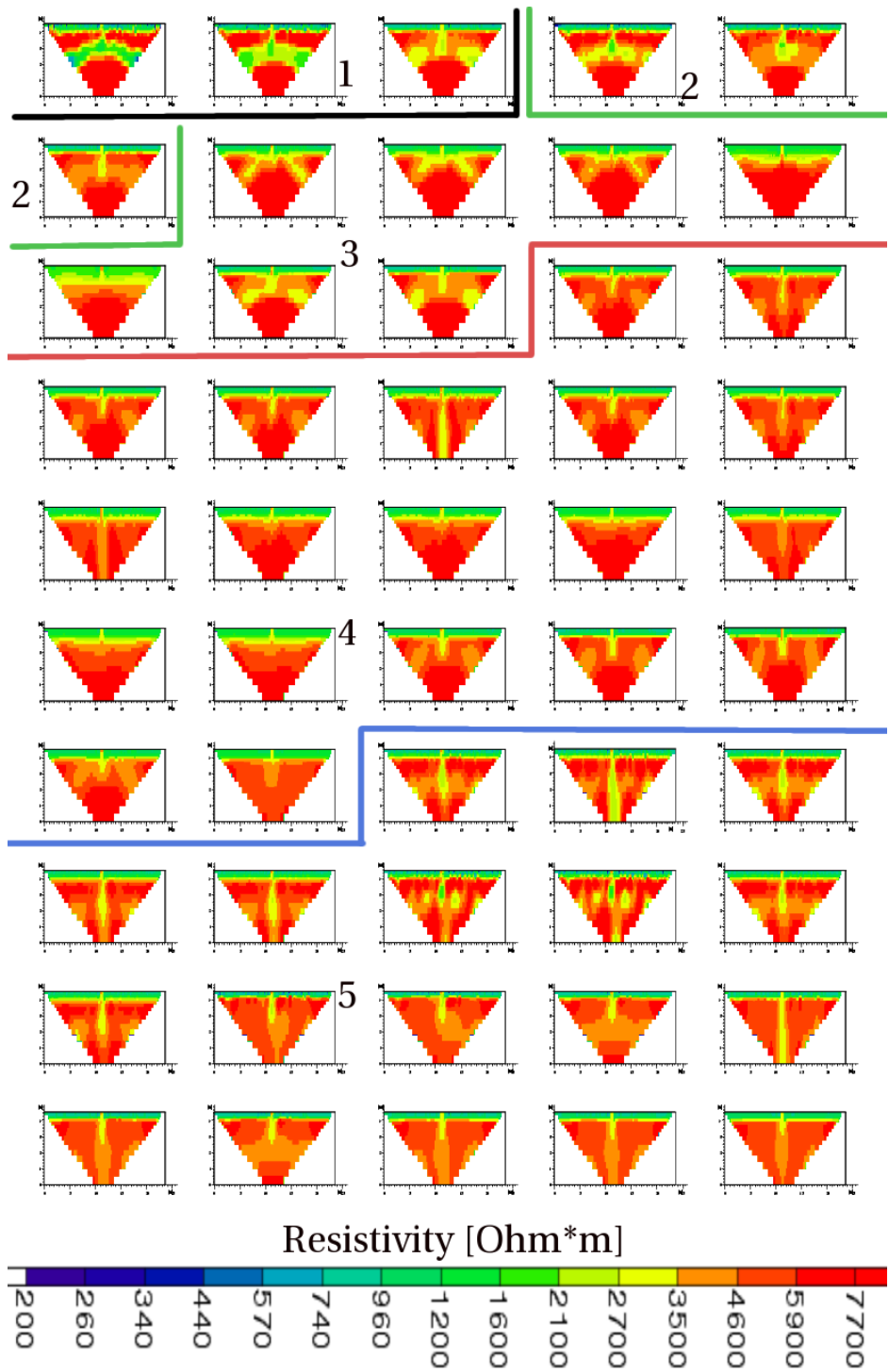}
\end{center}
\caption{\label{fig:ensemble}Clustered ensemble with 50 possible models for the diffusion state of the case of the defective sealing with an infiltration plume. The domains of the 5 clusters are indicated by numbers and dividing lines.}
\end{figure}

\begin{figure}
\begin{center}
\includegraphics*[width=.95\linewidth]{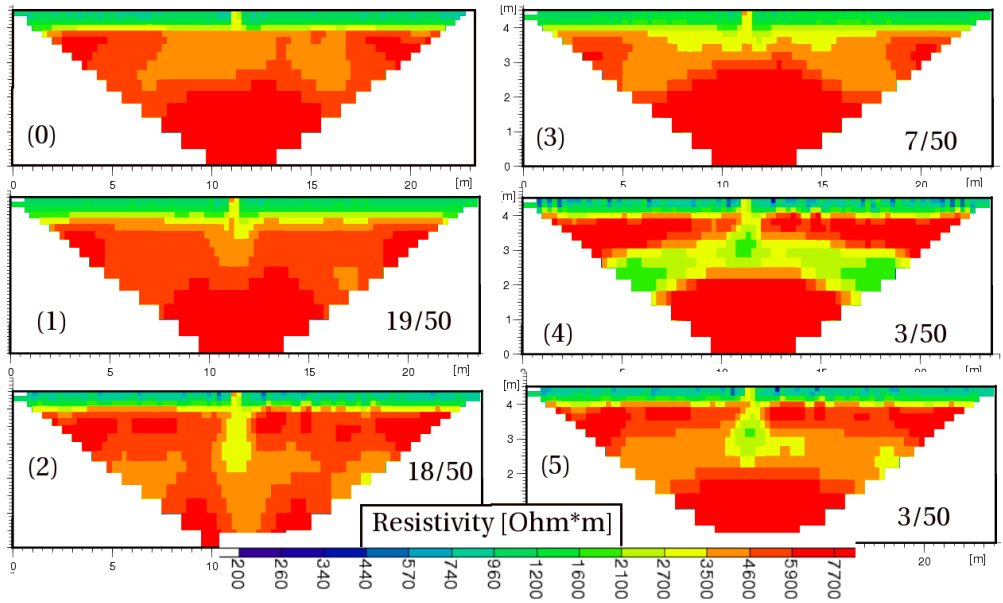}
\end{center}
\caption{\label{fig:merged-seal}Standard model (0) and averaged cluster models (1-5) for the case of the defective sealing. In contrast to Figure \ref{fig:ensemble}, the clusters are sorted in descending order by the number of ensemble members.}
\end{figure}

The inversion ensemble for the case of the defective sealing and the diffusion state is shown in Figure \ref{fig:ensemble}. All models within the ensemble detected the overburden with the damaged sealing, but the model parts below this overburden show different features. In the first cluster, $\Omega$-sloped artifacts appear to the side of the plume with equal resistivity as the plume itself. In the second cluster the artifacts appear as well, but have comparably higher resistivity, so that the plume appears as a distinct feature. In the third cluster, both plume and $\Omega$-sloped artifacts are roughly in the same resistivity range, but have a higher resistivity than in cluster 1. The fourth cluster comprises strongly damped models where the plume is mostly visible. The last cluster shows models where the plume is clearly visible, with comparably better contrast, but mostly the vertical extent of the plume feature is overestimated.\\
To comprehend the ensemble results in a simple way, averaged models of each cluster are shown in Figure \ref{fig:merged-seal}. As the clustering process already involves averaging, this is a valid method. In Figure \ref{fig:merged-seal}, the mean models for each of the clusters of the ensemble shown in Figure \ref{fig:ensemble} are now listed according to the number of cluster members. It must be noted that the smallest cluster contains only 3 models, whereas the largest cluster contains almost half the models of the ensemble. The average RMS error of each cluster is below 4\%.\\
The most prominent feature retrieved in all models is the two-layered structure, which can be observed in all five clusters. This structure is present even in clusters where the damping is strong enough to nearly hide the plume anomaly. When comparing clusters 3-5 to the strongly damped inversion results in cluster 1, the typical $\Omega$-sloped structure can be identified as an artifact at the lateral boundaries of the plume. Compared to the standard model (0), the cluster averages allow a much better identification of features, even though some interpretational experience or a priori knowledge is needed to distinguish between real anomalies (cluster 5) and artifacts (cluster 4).
\subsection{Second Case: Hydraulically Resistive Anomaly}
In the second case, the accuracy of resistivity quantification for a rectangular, hydraulically resistive anomaly placed below the organic overburden is studied. First, a soil model with an organic overburden and an anomaly at 0.55 m depth was created in HYDRUS. To represent the hydraulically resistive material of the anomaly, the same material as for the organic overburden was used. Then, multiple versions of this model were created with slightly different geometries. Table \ref{tab:possible} shows the differences between the respective models, which will be explained in the following.\\
\subsubsection{Water Simulation}
\begin{figure}
\begin{center}
\includegraphics*[width=.9\linewidth]{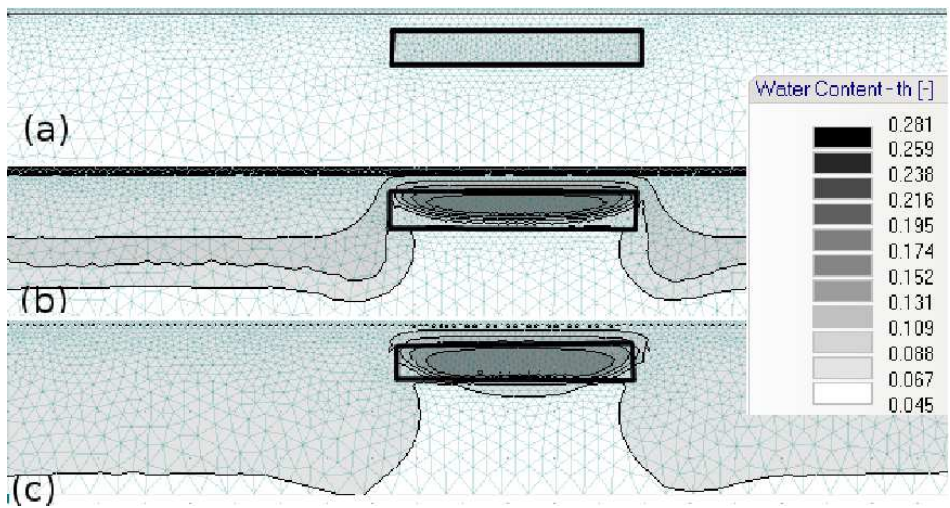}
\end{center}
\caption{\label{fig:block-hy}States of the simulation of water movement through a model with a hydraulically resistive anomaly (rectangular block marked with thick black outline). The layer boundary between organic overburden and sand is marked with a thin horizontal line. (a) Dry State (b) Infiltration State (c) Diffusion State.}
\end{figure}

In the simulation of water movement, a dry state, an infiltration state and a diffusion state were identified as characteristic states of an infiltrating water front. In the dry state (Fig. \ref{fig:block-hy}a), the soil is completely free of water. In the infiltration state (Fig. \ref{fig:block-hy}b), the water front is propagating into the volume. The hydraulically resistive anomaly causes water to impound on top, only slowly infiltrating into the anomaly. In the diffusion state (Fig. \ref{fig:block-hy}c), the infiltration front has reached the bottom boundary of the model, and the organic overburden and parts of the sand directly below are beginning to dry. The anomaly is filled with water that infiltrates into the sand beneath.\\
Analysis of the quality of water content estimation through ERT was conducted for a variety of models and electrode configurations based on the three states of water percolation in Figure \ref{fig:block-hy}. To study the influence of contrasting resistivities at the surface, models with and without an organic overburden were used for simulation. In addition, the depth of the anomaly was varied in steps of 0.2 m with the upper boundary at 0.35 m to 1.15 m depth. To examine the effect of electrode configuration, two different electrode arrays (complete Wenner-Schlumberger and Dipole-Dipole arrays) with an electrode spacing of 0.5 m were used for each model (Table \ref{tab:possible}).\\
\subsubsection{Forward-Inverse Cycle}
\begin{figure*}[tbp]
\begin{center}
\includegraphics*[width=\linewidth]{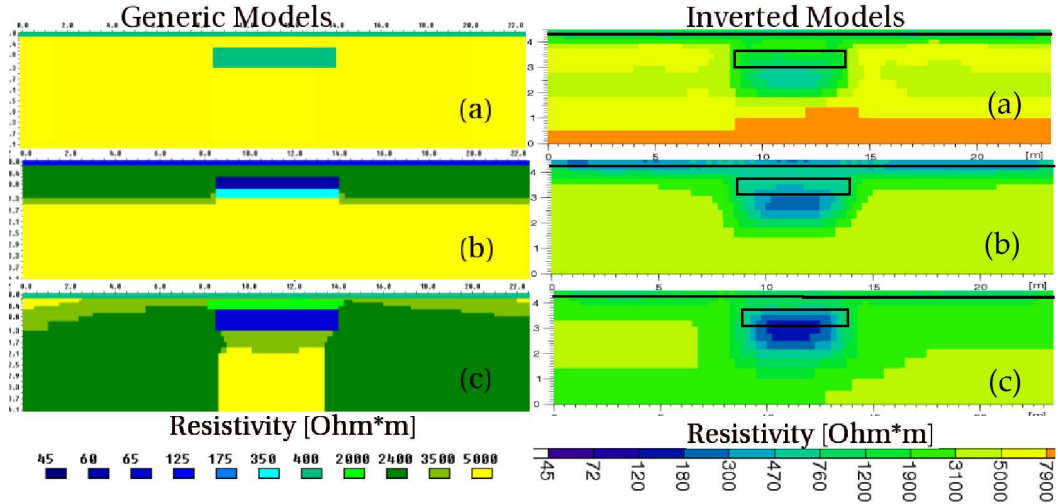}
\end{center}
               \caption{\label{fig:blocks}Generic and inverted models for the anomaly (Wenner-Schlumberger array). (a) Dry State (b) Infiltration State (c) Diffusion State. The black rectangle in the right column marks the location of the anomaly in the left column, the thin horizontal line marks the layer boundary between organic overburden and sand.}
\end{figure*}

\begin{figure}
\begin{center}
\includegraphics*[width=\linewidth]{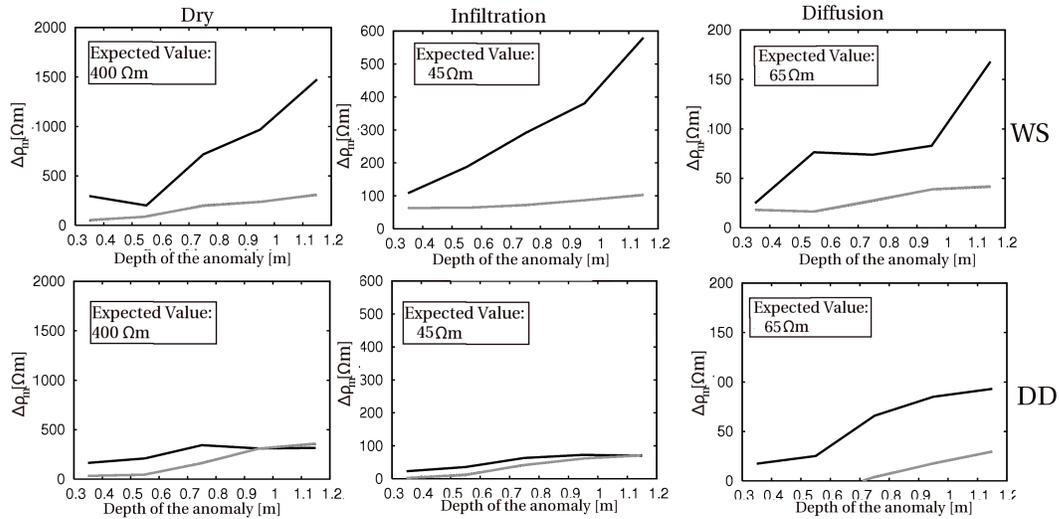}
\end{center}
\caption{\label{fig:misfit-plots}$\Delta\rho_m$ for cases with organic overburden (black lines) and without (gray lines). Top row: Survey with Wenner-Schlumberger, bottom row: with Dipole-Dipole. The left column shows the dry state, the middle column the infiltration state and the right column the diffusion state.}
\end{figure}

\begin{figure}
\begin{center}
\includegraphics*[width=8cm]{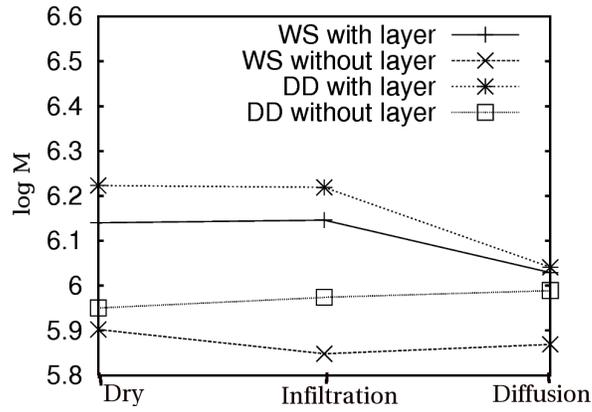}
\end{center}
\caption{\label{fig:cum-misfit}Cumulative block misfits for the three stages of water percolation. Shown is the logarithm of the sum of all errors $M$ for Wenner-Schlumberger and Dipole-Dipole arrays and with or without an organic overburden.}
\end{figure}

\begin{figure}
\begin{center}
\includegraphics*[width=7.7cm]{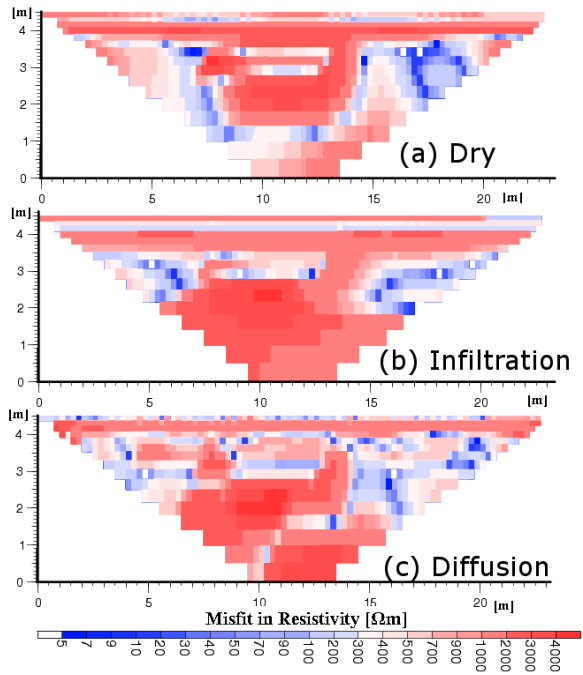}
\end{center}
\caption{\label{fig:mis-dis}Misfit in resistivity distribution by model blocks for anomaly at 0.95 m depth with organic overburden and Wenner-Schlumberger array.}
\end{figure}

Inspection of the inverted models (Fig. \ref{fig:blocks}, right column) shows that the rectangular shape of the anomaly cannot be exactly retrieved. Determination of an average resistivity of the anomaly would be dependent on an arbitrary determination of anomaly borders. It is also not possible to determine the average resistivity at the actual position of the anomaly, since the perceived depth of the anomaly is greater than the actual depth.\\ 
In the following, results for the different models shown in Table \ref{tab:possible} will be compared regarding $\Delta\rho_m$ (Eq. \ref{eq:misfit}), which now corresponds to the (minimal) resistivity of the anomaly. Figure \ref{fig:misfit-plots} shows $\Delta\rho_m$ as a function of anomaly depth.\\
For the Wenner-Schlumberger array, $\Delta\rho_m$ increases with anomaly depth, reaching up to 2-3 times the expected value. A much better estimate is obtained if no organic overburden is present (gray curves). For these cases, better quantifications of $\rho_{anom}^{-}$ are possible and $\Delta\rho_m$ increases only slightly with depth. In the diffusion state, significantly smaller errors occur compared to other states of water percolation, especially in the presence of an organic overburden.\\
As can be seen in Figure \ref{fig:blocks}, the error in depth resolution is rather large. If an organic overburden is present, the thickness of this layer is overestimated, causing a shift in the vertical position of the anomaly of 0.3 to 0.4 m. It was also observed that at greater depths, the position stays approximately the same for an anomaly expected at 0.75 m to 1.15 m depth. Again, in the case of a model without an organic overburden, the higher sensitivity due to higher resistivities near the surface makes better depth determination possible.\\
For models simulated with the Dipole-Dipole array, errors for models with organic overburden are significantly smaller than for the Wenner-Schlumberger array. However, the Dipole-Dipole array was shown to be very sensitive to noise and disturbances at the surface (like a stone pathway), to a point were measurements taken using this array could not be interpreted with the available inversion routines.\\
As a measure of the quality of the inversion, a simple criterion containing the model misfit $M$ as the sum of all errors has been applied:
\begin{equation}
M =\sum_{F_i} | \rho_{inv, i}-\rho_{gen, i} | 
\end{equation}

where $F_i$ is the $i$-th model block of the inversion domain discretisation.\\
Comparison of $M$ for the different states of water percolation (Fig. \ref{fig:cum-misfit}) shows that the diffusion state gives significantly better results. In this state, the misfit below the organic overburden and to the sides of the anomaly is much smaller, additionally the depth of the anomaly and overburden are better resolved.\\
Figure \ref{fig:mis-dis} shows the spatial error distribution for each state. In the dry state, the biggest errors stem from an overestimated thickness of the overburden, which also entails further mispositioning of the anomaly. The anomaly itself is also vertically elongated, leading to considerable errors in the lower parts. In the diffusion state (Fig. \ref{fig:mis-dis} (b)), the resistivity contrast between overburden and wet sand is much smaller, due to a) the sand having a reduced resistivity as it is more saturated with water and b) the overburden being dryer as in the previous states, resulting in a higher resistivity. As a consequence of this reduced resistivity contrast, the errors resulting from an incorrect overburden thickness are reduced as well.\\
\subsubsection{Ensemble}
\begin{figure}
\begin{center}
\includegraphics*[width=\linewidth]{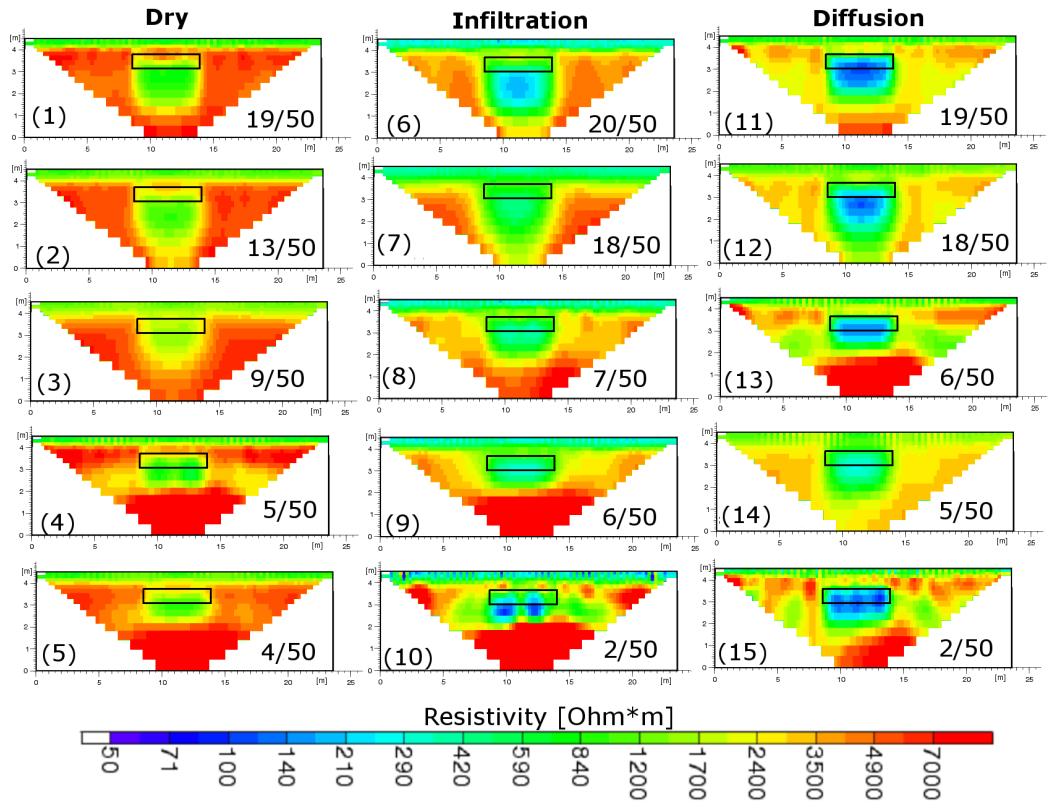}
\end{center}
\caption{\label{fig:aver-clus}Averaged cluster representatives for the resistive anomaly in the dry (left), infiltration (middle) and diffusion (right) state.}
\end{figure}

For the case of the hydraulically resistive anomaly, the random set of parameters is applied to generic models of all three different states of water percolation. To assure comparability, the random parameter set stays the same for each of the three models.\\
A model with an anomaly at 0.75 m depth was used, including an organic overburden and using the Wenner-Schlumberger array. For each state, an ensemble of 50 inverted models was created. For simplification, only mean cluster members are shown. Figure \ref{fig:aver-clus} shows the five clusters per ensemble with the respective number of ensemble members. The respective $\Delta\rho_m$ is listed in Table \ref{tab:misfit}. \\

\begin{itemize}
\item Dry State: The rectangular shape of the anomaly is retrieved variably well, but for the models 4 and 5, where the thickness of the anomaly is smaller, a strong overestimation of resistivities is present in the lower part of the model ($>7000~\Omega m$ instead of $5000~\Omega m$). The resistivity of the anomaly $\rho_{anom}^-$ is much too high for all five models. For models 1 and 2 that contain most of the ensemble members, the anomaly is vertically elongated.
\item Infiltration State: In four models, the shape of the anomaly has been retrieved quite well, but for model 10, two zones of minimal resistivity have been detected rather than the rectangular shape. In all models, $\Delta\rho_m$ is very large compared to the expected resistivity of the anomaly $\rho=65~\Omega m$. Again, models 9 and 10 (same inversion parameters as model 4 and 5) overestimate the background resistivity at greater depth.
\item Diffusion State: The resistivity of the anomaly is detected with lower resistivity as in the infiltration state, closer to the expected resistivity of $45~\Omega m$. Again, in model 13 and 15, the strong inversion artifact is present near the bottom coinciding with the shape of the anomaly being retrieved quite well. These artifacts are not present in model 11, 12 and 14, where the anomaly is vertically elongated. Model 15 presents a mixed case of a slightly elongated anomaly and an artifact of smaller extent than in model 13.
\end{itemize}
Table \ref{tab:misfit} shows, sorted for the cluster representatives, the misfits in the anomaly's resistivity. While it is apparent that the errors are large in each case, they are again considerably smaller for the diffusion state. 
\section{Discussion and Conclusion}
The ability of electrical resistivity tomography to accurately determine resistivity distributions was examined. A two-step model approach was used to create synthetic data sets. It comprises the modeling of soil water movement for synthetic soil data sets and a transfer into a model of generic resistivities using a petrophysical relation. A forward-inverse cycle is used evaluate how well the geophysical inversion scheme can reconstruct the given soil data set and its water content. An ensemble and clustering approach is proposed because a single model deduced as the optimal model does not necessarily reproduce the expected resistivities accurately.\\
The methods were applied to two case studies of simple soil models based on a two-layered structure reproducing field observations. The first case simulates the infiltration of water through a cracked surfical sealing, and the second a hydraulically resistive anomaly in a sand layer.\\
Key results of the forward-inverse modeling in this study include:
\begin{itemize}
\item In the presence of large resistivity contrasts, e.g. a conductive organic overburden, the retrieval of accurate resistivity values beneath this layer using the regularisation based inversion method applied in this study is not possible. However, if the volume is monitored at various stages of water percolation, the retrieval quality can differ. Especially in the diffusion state, much better accuracy was possible.
\item The model misfit increases with depth, as the sensitivity of the inversion model to the data decreases.
\item In the absence of an organic overburden, a much better quantification is possible because of a lower resistivity contrast.
\item The numerical study showed that a Dipole-Dipole array provides more accurate inversions than the Wenner-Schlumberger array. However, in practical applications, it has to be ensured that the signal-to-noise ratio is sufficiently large.
\end{itemize}
As a consequence, an ensemble approach was introduced that creates multiple inversion models for one data set by randomly choosing the inversion parameters from the possible (and numerically plausible) parameter space. By using clustering methods, averaged models representing different clusters in the ensemble can be created and compared. Key results of the ensemble approach include:\\
\begin{itemize}
\item Clustering of ensemble members allows an evaluation of the different possible models that fit the data. Areas likely to be plagued by artifacts can be identified and the reliability of standard inverted models can be evaluated.
\item However, the quantification of resistivities is not considerably improved by ensembles. For example, it became apparent that resistivities retrieved with smaller misfits in one region can coincide with larger artifacts in other regions.
\item The clustering of ensembles allows an overview of the ensemble, without losing information about the ensemble.
\end{itemize}
The ideas of the approaches presented here can easily be adapted to different models and inversion methods. For the specific inversion process with regularisation used in this study, it can be concluded that a reliable quantification of resistivity values is not possible. The use of additional information, e.g. within a framework aiming at directly inverting or calculating hydrological properties from collected data sets that not only contain resistivity measurements, but also data about the flow conditions, e.g. meteorological data, should be considered. 

\subsection*{Acknowledgments}
The authors thank M. Chouteau and one anonymous reviewer for their constructive comments which largely improved the manuscript. J. Rings acknowledges a grant from the Deutsche Forschungsgemeinschaft in the Postgraduate Programme Natural Disasters (GK 450).

\clearpage


\bibliographystyle{elsart-harv}
\bibliography{generell}

\begin{table}
\begin{center}
\begin{tabular}{|c||c|}
\hline
Constraint on the data&robust or smooth\\\hline
Constraint on the model&robust or smooth\\\hline
Initial damping $\lambda_i$&0.01 to 1\\\hline
Minimal damping $\lambda_m$&0.05$\lambda_i$ to 0.2$\lambda_i$\\\hline
Convergence limit&1\% to 9\%\\\hline
Maximal number of iterations&3, 5 or 15\\\hline
Vertical to horizontal regularisation&0.25 to 4\\\hline
Increase of damping with depth&1.0 to 2.0\\\hline
Reduce effect of &none, slight, \\
side blocks&severe, very severe\\\hline
Higher damping for first layer&yes or no\\\hline
\end{tabular}
\end{center}
\caption[Parameter space for ensembles]{\label{tab:ensemble}Parameter space of inversion parameters used for ensemble calculations.}
\end{table}
\begin{table}[tbp]
\begin{center}
\begin{tabular}{|c||c|c|c|c|c|}
\hline
Material&$\theta_r$&$\theta_S$&$\alpha$&$n$&$K_S$ $[m/d]$\\
\hline
Sand&0.045&0.361&4&2.2&17.28\\
Overburden&0.067&0.45&5.23&2.67&0.225\\      
\hline
\end{tabular}
\end{center}
\caption{\label{tab:materials}Soil parameters for the van Genuchten-Mualem parameterisation. $\theta_r$ is the residual water content, $\theta_S$ is the volumetric water content at full saturation, $\alpha$ and $n$ are parameters connected to the pore radii, and $K_s$ is the hydraulic conductivity at saturation.}
\end{table}

\begin{table}[tbp]
\begin{center}
\begin{tabular}{|c||c|c|c|c|c|}
\hline
Organic &with&without&&&\\
overburden&&&&&\\
\hline
Depth of &0.35&0.55&0.75&0.95&1.15\\
the anomaly [m]&&&&&\\
\hline
State of percolation&dry&infiltration&diffusion&&\\
\hline
ERT array&WS&DD&&&\\
\hline
\end{tabular}
\end{center}
\caption{\label{tab:possible}Parameter variation for different soil models, infiltration state and measurement geometry}
\end{table}

\begin{table}
\begin{center}
\begin{tabular}{|c|c|c|}
\hline
Cluster DRY & Cluster INFILTRATION & Cluster DIFFUSION\\
\hline
1) 381 & 6) 156 & 11) 49\\
2) 613 & 7) 294 & 12) 41\\
3) 228 & 8) 314 & 13) 86\\
4) 1068& 9) 73 & 14) 62\\
5) 608& 10) 292 & 15) 228\\
\hline
\end{tabular}
\end{center}
\caption{\label{tab:misfit}Misfit for the cluster representative shown in Figure \ref{fig:aver-clus} (misfits in $\Omega m$).}
\end{table}

\begin{figure}
\begin{center}
\includegraphics[width=.89\linewidth]{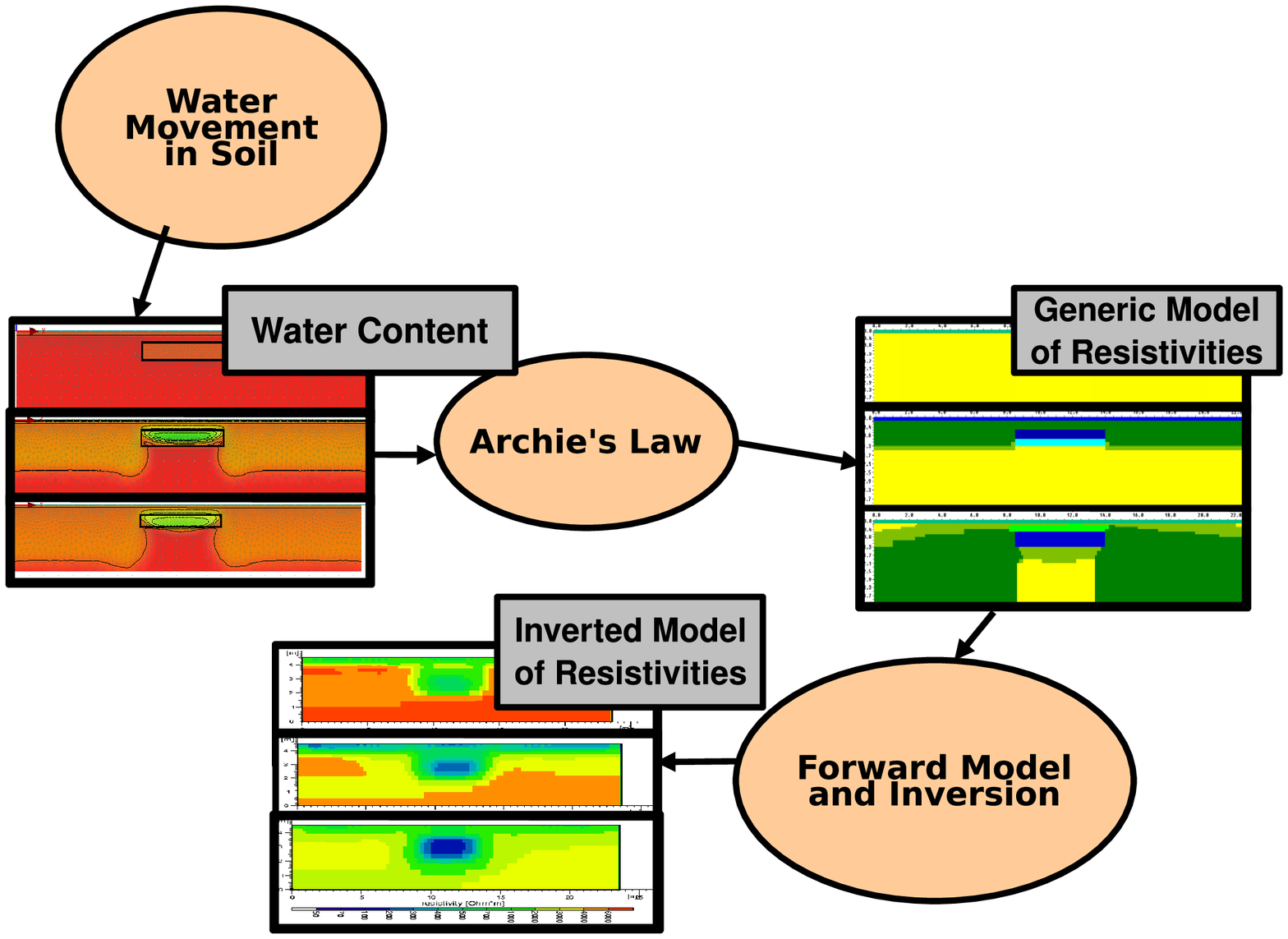}\\
\includegraphics[width=.89\linewidth]{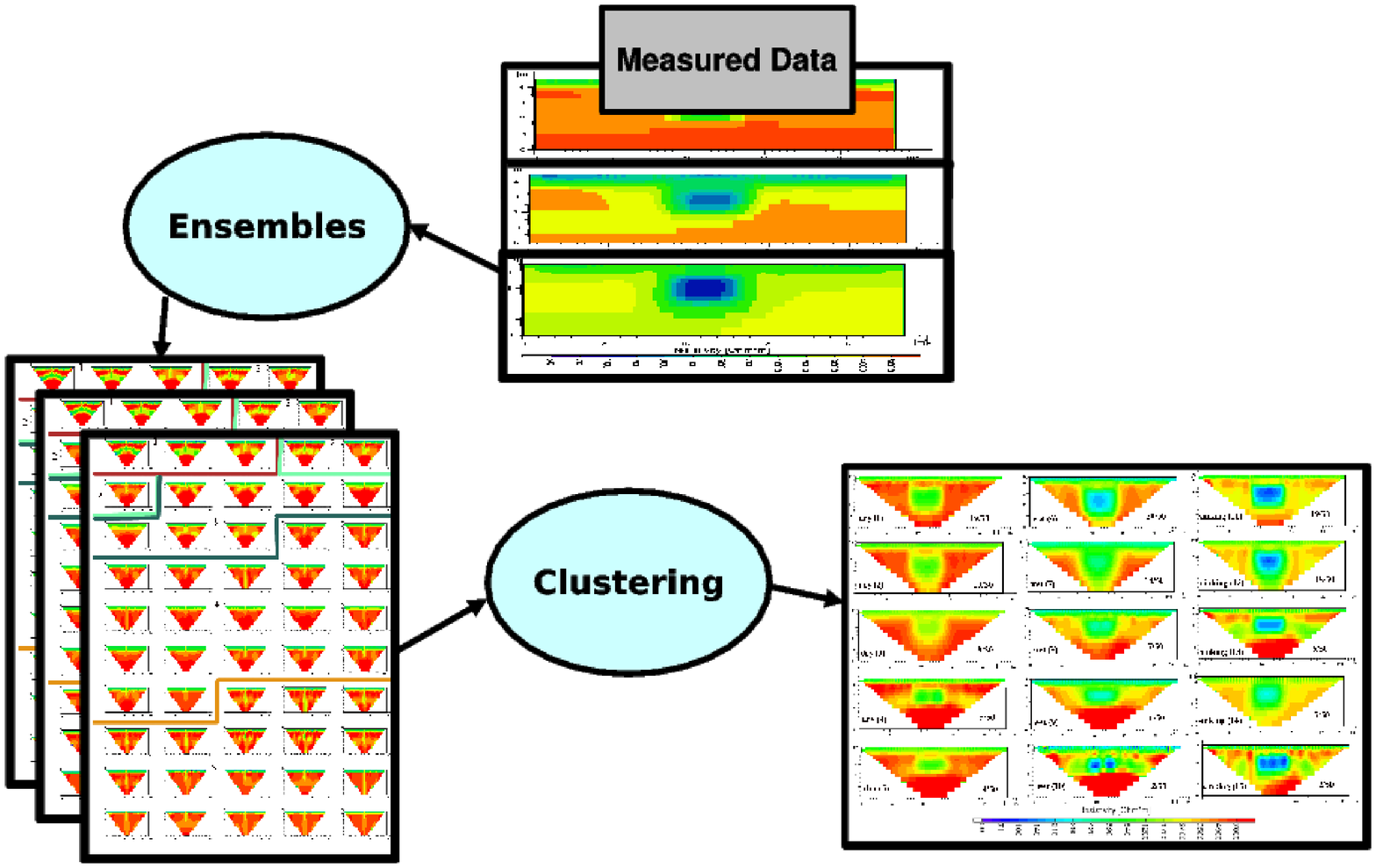}
\end{center}
\caption{\label{fig:chart}Charts visualizing the methodological steps involved in this study. Above: Steps in the forward-inverse cycle. Below: Steps in the ensemble method.}
\end{figure}

\clearpage

\end{document}